\documentstyle[aps]{revtex}

\begin{document}
\title{Permanent magnetic moment in mesoscopic metals with spin-orbit interaction}
\author{R.A. Serota\thanks{%
serota@physics.uc.edu}}
\maketitle

\begin{abstract}
We argue that at zero temperature an isolated metal particle (or an AB\
ring) with spin-orbit interaction and odd number of electrons will have a
permanent magnetic moment, even in zero magnetic field (flux). In a
zero-field-cooled state both the direction and the magnitude of the moment
varies from particle to particle and averages to zero. In a field-cooled
state it averages to $\sim \mu _{B}\left( k_{F}\ell \right) ^{1/2}$. We
argue that the permanent moment is due to an uncompensated electron in the
last occupied (Fermi) level. We introduce an effective single-electron
Hamiltonian which accounts for spin-orbit\ coupling.
\end{abstract}

\section{Introduction}

Permanent currents in mesoscopic SNS junctions, that is currents flowing in
the absence of the phase difference between the superconductors, were
predicted by Altshuler and Spivak\cite{AS} in the circumstance of a
time-reversal symmetry breaking perturbation, such as spin-flip scattering.
Since their model is equivalent to that of an Aharonov-Bohm (AB) ring
threaded by a flux, it also implies a permanent AB\ (persistent) current
(or, in general, a permanent magnetic moment) when the time-reversal
symmetry is broken. Such currents (moments), however, average out to zero
and are uncorrelated with the flux (field).

Kravtsov and Zirnbauer (KZ)\cite{KZ} predicted permanent AB currents in
rings with odd numbers of electrons and (non time-reversal breaking)
spin-orbit (SO) scattering. Their argument was based on the Kramers' theorem%
\cite{LL} in the context of disordered mesoscopic systems\cite{ASh}. Namely,
the level degeneracy should remain two-fold, where a state and a
time-reversed state are two different states at the same energy. KZ realized
that in the presence of SO coupling these states are current-carrying. To
estimate the current's magnitude for strong SO they used the level
correlation function of a symplectic ensemble\cite{M},\cite{E}. The letter
contains a peaked function of the flux\cite{MP},\cite{AIE} which, in the
case of a zero flux,\ reduces to a $\delta -$function. The physical
interpretation suggests that the double degeneracy is lifted by the flux and
one of the two current-carrying states becomes preferable; in zero flux
either state can be occupied with equal probability.

In the absence of SO scattering the two-fold Kramers' degeneracy for an odd
number of electrons is trivial in that it corresponds to the spin-up and
spin-down states of an uncompensated electron in the last occupied (Fermi)\
level (all other levels being doubly occupied). For a sufficiently strong
SO, spin is no longer a good quantum number and one must consider an
effective moment, whose origin is in mixing of orbital and spin degrees of
freedom and which is conserved\cite{LL} since a metal particle or a ring can
be viewed as a closed system. However, unlike atomic physics, for instance,
where the nature of such moment is obvious\cite{LL} as ${\bf J}={\bf L}+{\bf %
S}$, in a disordered conductor $L$ is not a good quantum number. It is
clear, on the other hand, that the moment is due to the last uncompensated
electron, as in spin magnetism.

Since one must differentiate between the odd and even numbers of electrons,
understanding of the nature of permanent moments requires formalism that
accounts for a single-electron contribution. The purpose of this work is to
introduce a single-electron Hamiltonian that would effectively incorporate
SO coupling.

In a previous paper\cite{SS}, we addressed a more rigorous derivation of the
permanent AB current in the SO case. Towards that end we applied a so called
''mixed'' approach where the Imry's formula\cite{I} 
\begin{eqnarray}
F-\Omega \left( \left\langle \mu \right\rangle \right) &=&\frac{1}{%
2\left\langle \upsilon \right\rangle }\left\langle \delta N^{2}\right\rangle
\label{F--Omega} \\
&=&\frac{1}{2\left\langle \upsilon \right\rangle }\int \int d\varepsilon
_{1}d\varepsilon _{2}\left\langle \delta \upsilon \left( \varepsilon
_{1},H\right) \delta \upsilon \left( \varepsilon _{2},H\right) \right\rangle
f\left( \varepsilon _{1},\left\langle \mu \right\rangle \right) f\left(
\varepsilon _{2},\left\langle \mu \right\rangle \right)
\label{F--Omega-expanded}
\end{eqnarray}
relating the free energies of the canonical and grand canonical ensembles%
\footnote{%
In a closed Kubo particle the total number of electrons is presumed fixed,
hence one must derive the free energy of a canonical ensemble\cite{K}.} is
used in combination with the exact formulae\cite{MP},\cite{AIE} for the
disorder-averaged level-density correlation function $\left\langle \delta
\upsilon _{1}\delta \upsilon _{2}\right\rangle $. We call this approach
''mixed'' because (\ref{F--Omega}) implies large particle number
fluctuations in the equivalent grand canonical ensemble, $\left\langle
\delta N^{2}\right\rangle \gg 1$, and consequently\cite{LL2} $T\gg \Delta $,
where $\Delta $ is the mean energy-level spacing. In such a circumstance a
perturbative expression\cite{ASh} for the level correlation function should,
in principle, be used. However, as will be discussed below, the ''mixed''
approach gives qualitatively correct results on approaching the
level-quantized limit for $T\sim \Delta $, when the single-electron effects
become important.

In what follows, we will first compare the results obtained from (\ref
{F--Omega}) in the ''mixed'' approach for spin and orbital magnetism with
those obtained in the perturbative limit and\ in a single-electron
approximation in the level quantized limit\cite{SS},\cite{OZS}-\cite{S}.
Such comparison will enable us to gain insight into the level-quantized
problem in the SO\ case by inference from the ''mixed'' approximation. We
proceed to introduce a single-electron Hamiltonian which effectively
includes SO\ coupling. We discuss the physical consequences of our ansatz
and check them against other known results, such as mesoscopic fluctuations.

\section{Perturbative, ''mixed'' and single-electron approximations}

The results in this Section are derived in our previous papers\cite{SS},\cite
{OZS}-\cite{S}. For simplicity, we consider only 2D\ systems. We begin with
the spin magnetic susceptibility and use the notations 
\begin{equation}
\chi _{P}=\mu _{B}^{2}\left\langle \upsilon \right\rangle \text{ and }\chi
_{C}=\frac{\mu _{B}^{2}}{T}  \label{khi_Curie-khi_Pauli}
\end{equation}
for the Pauli and Curie susceptibility respectively, where $\left\langle
\upsilon \right\rangle $ is the mean density of levels at the Fermi surface
and $\left\langle \upsilon \right\rangle \Delta =s=2$ is the level
degeneracy. The summary for the mean susceptibility is given in Table I:

\begin{center}
TABLE\ I:\ Spin magnetic susceptibility
\end{center}

\begin{equation}
\begin{tabular}{|c|c|c|}
\hline
$\text{Pertubative}$ & $\text{''Mixed''}$ & $\text{Single-electron}$ \\ 
\hline
$T\gg \Delta $ & $T\sim \Delta $ & $T\ll \Delta $ \\ \hline
$\chi _{P}\left[ 1+\frac{3\zeta \left( 3\right) }{\pi ^{4}}\left( \frac{%
\Delta }{T}\right) ^{2}\right] $ & $\frac{2}{3}\chi _{C}$ & 
\begin{tabular}{cc}
$\chi _{C}$ & $\text{od}$d number of electrons \\ 
$const\left[ \chi _{P}\left( \frac{T}{\Delta }\right) ^{\alpha }\right] $ & $%
\text{even}$ number of electrons
\end{tabular}
\\ \hline
\end{tabular}
\label{khi-all}
\end{equation}
Clearly, as the temperature approaches the level spacing, the ''mixed''
approach correctly predicts the onset of the Curie-like susceptibility.
Conversely, it is beyond its range of applicability to distinguish between
the odd and even numbers of electrons and give a single-electron
contribution. (In a sense, it averages between the odd and even
contributions which is dominated by the odd-case Curie susceptibility.) For
further comparison with the SO\ case, we note that the ''mixed'' result is
due to Zeeman splitting of the $\delta -$function in the exact
level-correlation function. We also note that the linear response regime for
the spin susceptibility extends to the fields $\mu _{B}H\sim \Delta $, at
which point the energy levels begin to separate into two non-interacting
series of levels with spin-up and spin-down electrons\cite{ASh} and the
magnetic component of the free energy (that is, the change in free energy
due to Zeeman splitting) approaches $-\Delta $.

(We briefly discuss the case of even number of electrons\cite{DMS}. The
constant and the exponent $\alpha $ in Table I depend on the symmetry\cite{M}%
: $\alpha =1$ for Gaussian orthogonal ensemble (GOE) and $\alpha =2$ for
Gaussian unitary ensemble (GUE). Notice that GUE\ was not considered in\cite
{DMS} since it was assumed that the transition from GOE\ to GUE\ takes place
at $\mu _{B}H\sim \Delta $ when, in reality, it takes place at $\mu
_{B}H\sim \Delta \sqrt{\Delta /E_{c}}\ll \Delta $ (see eq. (\ref{phi_c})
below and Refs.\cite{E},\cite{ASh}). On the other hand, for a strong SO, the
spin susceptibility in the Gaussian symplectic ensemble (GSE),\ considered in%
\cite{DMS}, is not particularly meaningful since spin is not a good quantum
number in this case).

We next turn to orbital magnetism in the absence of SO\ interaction. This
case is summarized in Table II in terms of the mean AB (persistent) current $%
I\left( \phi \right) $:

\begin{center}
TABLE\ II:\ Persistent current in the absence of SO
\end{center}

\begin{equation}
\begin{tabular}{|c|c|c|}
\hline
$\text{Pertubative}$ & $\text{''Mixed''}$ & $\text{Single-electron}$ \\ 
\hline
$T\gg \Delta $ & $T\sim \Delta $ & $T\ll \Delta $ \\ \hline
\begin{tabular}{cc}
$\frac{4}{3}eE_{c}\frac{\Delta }{T}\frac{\phi }{\phi _{0}}$ & $\frac{1}{2\pi
^{3}}e\Delta \frac{\phi _{0}}{\phi }$ \\ 
$\phi \ll \phi _{c}$ & $\phi _{c}\ll \phi \ll \phi _{0}$%
\end{tabular}
& 
\begin{tabular}{cc}
$8eE_{c}\frac{\phi }{\phi _{0}}$ & $\frac{1}{2\pi ^{3}}e\Delta \frac{\phi
_{0}}{\phi }$ \\ 
$\phi \ll \phi _{c}$ & $\phi _{c}\ll \phi \ll \phi _{0}$%
\end{tabular}
& 
\begin{tabular}{c}
$\sim 2\pi eE_{c}\frac{\phi }{\phi _{0}}$ \\ 
$\phi \ll \phi _{c}$%
\end{tabular}
\\ \hline
\end{tabular}
\label{I_orb-all}
\end{equation}
where $\phi _{0}=2\pi /e$ is the flux quantum (in the units where $\hbar
=c=1 $) and $E_{c}=D/L^{2}$ where $D$ is the diffusion coefficient and $L$
is the circumference of the ring. The flux scale $\phi _{c}$ defines the
limit for linear response and, at $T=0$, 
\begin{equation}
\phi _{c}\sim \phi _{0}\sqrt{\frac{\Delta }{E_{c}}}  \label{phi_c}
\end{equation}
sets the scale of the transition from GOE to GUE\cite{E}. We point out that
the ''mixed'' approach correctly predicts current saturation as $T$ crosses $%
\Delta $, which is confirmed by a single-electron type calculation\cite{S}
where the orbital magnetic susceptibility is attributed to the van Vleck
response involving virtual transitions between the last occupied and the
first unoccupied states at the Fermi level. We also note that, as $\phi $
crosses $\phi _{c}$, the maximal current is $\sim e\Delta \left( k_{F}\ell
\right) ^{1/2}$ (corresponding to a magnetic moment of order $\mu _{B}\left(
E_{c}/\Delta \right) ^{1/2}\sim \mu _{B}\left( k_{F}\ell \right) ^{1/2}$),
where $k_{F}$ is the Fermi wave vector and $\ell $ is the electron
mean-free-path, and the magnetic energy $-I\left( \phi \right) \phi $
saturates to $-\Delta $.

We finally turn to the SO case. The results are summarized in Table III:

\begin{center}
TABLE\ III:\ Persistent current with strong SO
\end{center}

\begin{equation}
\begin{tabular}{|c|c|c|}
\hline
$\text{Pertubative}$ & $\text{''Mixed''}$ & $\text{Single-electron}$ \\ 
\hline
$T\gg \Delta $ & $T\sim \Delta $ & $T\ll \Delta $ \\ \hline
\begin{tabular}{cc}
$\frac{1}{3}eE_{c}\frac{\Delta }{T}\frac{\phi }{\phi _{0}}$ & $\frac{1}{8\pi
^{3}}e\Delta \frac{\phi _{0}}{\phi }$ \\ 
$\phi \ll \phi _{c}$ & $\phi _{c}\ll \phi \ll \phi _{0}$%
\end{tabular}
& 
\begin{tabular}{cc}
$eE_{c}\frac{\Delta }{T}\frac{\phi }{\phi _{0}}$ & $\frac{1}{8\pi ^{3}}%
e\Delta \frac{\phi _{0}}{\phi }$ \\ 
$\phi \ll \phi _{c}$ & $\phi _{c}\ll \phi \ll \phi _{0}$%
\end{tabular}
& 
\begin{tabular}{c}
$\stackrel{?}{\sim }eE_{c}\frac{\Delta }{T}\frac{\phi }{\phi _{0}}$ \\ 
$\phi \ll \phi _{c}$%
\end{tabular}
\\ \hline
\end{tabular}
\label{I_sym-all}
\end{equation}
where the question mark indicates the result inferred from the ''mixed''
approach. A\ formal extension to $T=0$ in the mixed approach yields a
permanent current for $\phi =0$\cite{SS}, 
\begin{equation}
I\left( 0\right) =\frac{e}{2\pi }\sqrt{2E_{c}\Delta }  \label{I_0}
\end{equation}
and the same order of magnitude for a permanent AB current is expected in
the single-electron picture (and a corresponding magnetic moment of order $%
\mu _{B}\left( E_{c}/\Delta \right) ^{1/2}\sim \mu _{B}\left( k_{F}\ell
\right) ^{1/2}$). In the next section we introduce a single-electron
Hamiltonian which indeed reproduces this result.

\section{Effective single-electron Hamiltonian}

A single-electron Hamiltonian which effectively accounts for multiple SO
scattering effects can be written as 
\begin{equation}
H_{eff}=\lambda {\bf L}\cdot {\bf S}+\mu _{B}{\bf H}\cdot \left( {\bf L}+2%
{\bf S}\right)   \label{H_eff}
\end{equation}
where 
\begin{equation}
\lambda \sim \Delta \sqrt{\frac{\Delta }{E_{c}}}\sim \frac{\Delta }{\left(
k_{F}\ell \right) ^{1/2}}  \label{lambda}
\end{equation}
In the 2D case considered here and magnetic field perpendicular to the plane
of the sample, one need to consider only $z-$components of the spin and
angular momentum. Treated as a perturbation to second order, the Hamiltonian
(\ref{H_eff}) yields the energy shift given by 
\begin{equation}
\delta E=-\mu _{B}2\left( 1+\lambda \Lambda _{zz}\right) S_{z}H-\mu
_{B}^{2}\Lambda _{zz}H^{2}-\lambda ^{2}\Lambda _{zz}S_{z}^{2}  \label{deltaE}
\end{equation}
where 
\begin{equation}
\Lambda _{zz}=\frac{\left| \left\langle i\right| L_{z}\left| f\right\rangle
\right| ^{2}}{\varepsilon _{f}-\varepsilon _{i}}  \label{Lambda_zz}
\end{equation}
The first term in (\ref{deltaE}) describes the coupling of the permanent
moment - a combination of spin and SO-induced components - to the magnetic
field. The second term is the van Vleck term due strictly to orbital
effects. The last term describes the single-axis anisotropy. In eq. (\ref
{Lambda_zz}) we limit our consideration to the virtual transitions between
the last occupied (Fermi) level $i$ and the first unoccupied level $f$, so
that $\Lambda _{zz}>0$.

In disordered conductors the orbital magnetic effects are large due to
diffusion\cite{AAZ}. In the present case, it is reflected in largeness of $%
\Lambda $ which was evaluated in \cite{S} 
\begin{equation}
\Lambda _{zz}\sim \frac{E_{c}}{\Delta ^{2}}\sim \frac{k_{F}\ell }{\Delta }
\label{Lambda_zz-est}
\end{equation}
Consequently for the magnetic moment we find 
\begin{equation}
M=-\frac{\partial \delta E}{\partial H}=const_{1}\left[ \mu _{B}\left(
k_{F}\ell \right) ^{1/2}\right] +const_{2}\left[ \left| \chi _{L}\right|
\left( k_{F}\ell \right) H\right]   \label{Moment}
\end{equation}
where $\chi _{L}$ is the Landau susceptibility\cite{LL2}. The first term in (%
\ref{Moment})\ is the permanent magnetic moment due to SO\ coupling and the
second term is the induced van Vleck orbital moment.

Independent checks of ansatz (\ref{H_eff}) come from the consideration of
the anisotropy term (the third term in (\ref{deltaE})) and the mesoscopic
fluctuations of spin-polarization\cite{ZS}. For the former, using eqs. (\ref
{lambda})\ and (\ref{Lambda_zz-est}), we find $\delta E\sim -\Delta $, which
is expected for the energy associated with the transition to GSE. For the
latter, the large fluctuations of local magnetization were interpreted in
terms of the enhancement of the effective electron $g-$factor, $g=2\left(
1+\lambda \Lambda _{zz}\right) $, to $\sim \left( k_{F}\ell \right) ^{1/2}$,
which is consistent with the present results. Another consistency check
comes from the requirement that both the linear and quadratic\ magnetic
terms in (\ref{deltaE}) be less than $\Delta $ and we find that either one
is equivalent to the condition $\phi <\phi _{c}$ for the transition to GUE,
where $\phi _{c}$ is given by eq. (\ref{phi_c}).

\section{Discussion}

We proposed the single-electron Hamiltonian (\ref{H_eff})\ which effectively
incorporates the effect of multiple SO scatterings for the electron in the
Fermi level. The resulting magnitude (\ref{Moment}) of the permanent
magnetic moment of a 2D ring or a simply-connected conductor is consistent
with the results based on the perturbative and ''mixed'' approaches. We also
reproduce the correct energy scales and the enhanced magnitude $\sim \left(
k_{F}\ell \right) ^{1/2}$ of the electron $g-$factor. However, at present
there is no microscopic derivation of (\ref{H_eff}). We hope to address this
in a future work.

\section{Acknowledgments}

I am grateful to Bernard Goodman for many useful discussions and reading of
this manuscript. This work was not supported by any funding agency.

{\it Note added}: After the completion of this paper I became aware of the
manuscript cond-mat/0001431 which discusses some of the same issues. For 2D
systems there is a general agreement between the two and with our previous
work\cite{ZS}.

\end{document}